\documentclass[runningheads]{llncs}
\usepackage[T1]{fontenc}
\usepackage{url}
\usepackage{graphicx}
\graphicspath{{figures/}}
\usepackage{subcaption}
\usepackage{comment}
\usepackage{booktabs}
\usepackage[inline]{enumitem}
\begin{document}
\title{A Task-Level Bias Taxonomy for Historical Image Retrieval}
%
%
\author{Omkar Gavali\inst{1}\orcidID{0009-0003-4735-6469} \and
Ratan Sebastian\inst{2,3}\orcidID{0000-0002-8034-3382} \and
Christoph Rippe\inst{4}\orcidID{0000-0003-3209-2504} \and
Ralph Ewerth\inst{1,5,2,3}\orcidID{0000-0003-0918-6297} \and
Anett Hoppe\inst{1,5,2}\orcidID{0000-0002-1452-9509}}
\authorrunning{Gavali et al.}
%

\institute{
Philipps-Universit\"at Marburg, Germany \and
TIB -- Leibniz Information Centre for Science and Technology, Hannover, Germany \and
L3S Research Center, Leibniz University Hannover, Germany \and
Goethe University Frankfurt, University Library, Germany \and
Hessian Center for Artificial Intelligence (hessian.AI), Marburg/Darmstadt, Germany}
\maketitle              

\begin{abstract}
In recent decades, galleries, libraries, archives, and museums~(GLAM institutions) have digitised vast amounts of cultural heritage material, including historical imagery. A major challenge in making these collections publicly accessible is enabling effective search and discovery. Many historical images contain incomplete, sparse, or no descriptive metadata. Consequently, machine learning~(ML) methods are increasingly being used to generate or enrich metadata to improve the searchability of such collections.
However, artificial intelligence~(AI)-based systems can reproduce or amplify existing biases, particularly when applied to historical imagery. Factors such as differences between the data used to train the models and the historical images on which they are applied, degraded image quality, monochrome photographs, and historically biased or incomplete descriptions can reduce model performance and introduce bias into search and retrieval results. In this paper, we examine different forms of bias in historical image collections and review current methods to mitigate them. We investigate how these biases arise and how they influence AI-supported search and retrieval systems.

\end{abstract}
\section{Introduction}
Historical image collections are not neutral: the categories used to describe them, and the omissions of what was documented, reflect the priorities and power relations of the period and institutions that produced them~\cite{douglas2018,BlochMartosSantana2025,king2024archival}.
These issues are particularly evident in colonial collections, which were shaped by unequal power relations and colonial perspectives. When such collections are made searchable, particularly where retrieval is supported with modern machine learning (ML) methods, these embedded assumptions are not simply preserved but risk being actively reproduced or reinforced, shaping how present-day users access and understand the past~\cite{foka2025,dentler2024sensitivity}. 

Bias and its mitigation have so far been predominantly studied for modern imagery, and existing mitigation techniques tend to work best for such images.
Comparatively little research addresses bias and bias mitigation in historical imagery, particularly where descriptive metadata is sparse or absent~\cite{foka2025,dentler2024sensitivity}.
In this paper, we address this gap, examining how bias arises in historical image collections and which mitigation techniques remain applicable.  

We ground this discussion in the case of colonial photograph archives, drawing throughout on the Visual Analytics for Images from Colonial Contexts~(VABiKo) project as a concrete instance of the challenges discussed here; though bias dimensions and mitigation gaps are not specific to this collection alone.
The VABiKo collection comprises more than 50,000 photographs collected between 1887 and 1943 by the German Colonial Society and its successor organizations, and held today by the Universitätsbibliothek J.\,C.\ Senckenberg (UB JCS) in Frankfurt am Main~\cite{ubfrankfurt_bildsammlung}.
As illustration of the particularities that this material raises, these images are predominantly monochrome and, having been digitized from glass-plate negatives and prints by 1990s microfilming, may be physically degraded. They can also depict sensitive content, including violence, nudity, or displays of colonial power. We return to these particularities and their consequences for bias in detail in Section~\ref{sec:vabiko_data}.

Bias itself is framed differently across disciplines.
In the humanities, it is typically understood as prejudice or unfair treatment directed at a group based on attributes such as race or gender~\cite{foka2025}.
In machine learning, it instead denotes a property introduced at specific stages of the data-algorithm-user pipeline, assessed by evaluating system outputs.
Effective mitigation, however, has been shown to depend on interdisciplinary collaboration between heritage professionals, subject-matter experts and data scientists~\cite{jaillant2025can}.
We therefore adopt a socio-technical framing that treats these two perspectives as complementary.


This paper makes two main contributions:
\begin{enumerate*}[label=(\arabic*) ]
 \item a systematic mapping of bias categories from an established ML fairness taxonomy onto retrieval tasks relevant to colonial-context photography archives, grounded in the VaBiKo collection, and
 \item an analysis of standard mitigation techniques and their limitations for these biases.
\end{enumerate*}

The remainder of this paper is structured as follows. Section~\ref{sec:related_work} reviews bias from humanities and machine-learning perspectives and surveys general-purpose bias mitigation techniques. Section~\ref{sec:context_vabiko} introduces the VaBiKo project, its collection and six retrieval tasks relevant to it, deriving the bias dimensions each raises. Section~\ref{sec:vabiko_bias_mitigation} examines mitigation techniques for these dimensions and where their assumptions hold or break for colonial-context material. Section~\ref{sec:conclusion} concludes with open problems for future work.

\section{Related Work}
\label{sec:related_work}
Understanding bias in AI-supported retrieval systems requires drawing on two distinct bodies of work: research that defines and categorises bias as a concept, and research that proposes concrete techniques to detect and mitigate it.
Section~\ref{sec:rw_bias_perspectives} reviews how bias has been conceptualised both in the humanities and in machine learning.
We outline which of these perspectives, and which pipelines stages are the most relevant to historical image collections.
Section~\ref{sec:rw_bias_mitigation} then surveys existing mitigation techniques developed primarily for contemporary imagery and datasets, providing the technical baseline against which we later assess applicability to the colonial-context material described in Section~\ref{sec:context_vabiko}.


\subsection{Bias Perspectives}
\label{sec:rw_bias_perspectives}
Bias in historical collections cannot be fully understood from a single disciplinary perspective; therefore, we draw on both the humanities and machine learning literature before deriving the dimensions most relevant to our own context as presented in Section~\ref{sec:context_vabiko}.

\paragraph{Bias in the Digital Humanities: }
Most historical collections contain some form of metadata, such as location, date, title, and description, often structured through ontologies or classification systems inherited from colonial perspectives.
These metadata may reflect displays of power, propaganda, and other colonial ways of representing people, places, and events.
From a humanities perspective, bias, therefore, begins with how historical material was originally produced, collected, selected, described, and curated over time~\cite{dentler2024sensitivity,king2024archival}.

In the case of historical imagery, this includes decisions about who had the authority to produce photographs, which people and activities were photographed, how subjects were represented, which images entered institutional collections, which materials survived, and which objects were later selected for digitisation~\cite{dentler2024sensitivity}.
These processes may create archival silences and uneven visibility, in which some communities, geographic regions, and historical perspectives are extensively documented, while others remain absent or difficult to access~\cite{BlochMartosSantana2025}.
In the present VaBiKo collection (Section~\ref{sec:vabiko_data}), these issues are visible in metadata fields such as historical ethnic ascriptions, content keywords, depicted persons, and locations, which may preserve externally imposed colonial categories, contain incomplete identifications, or omit the perspectives of the people represented.

When machine-learning methods are used to generate or enrich metadata, there is also a risk of digital violence if the same historical categories, terminologies, metadata structures, and ontologies are transferred into modern digital infrastructures.
In this way, colonial perspectives embedded in historical collections may not only be preserved but also reproduced and reinforced through AI-based systems~\cite{BlochMartosSantana2025}.
This humanities perspective therefore explains how bias becomes embedded in the source material and metadata before it is operationalised as data and processed by machine-learning systems~\cite{BlochMartosSantana2025,foka2025}.

\begin{figure}[htbp]
    \centering
    \includegraphics[width=\textwidth]{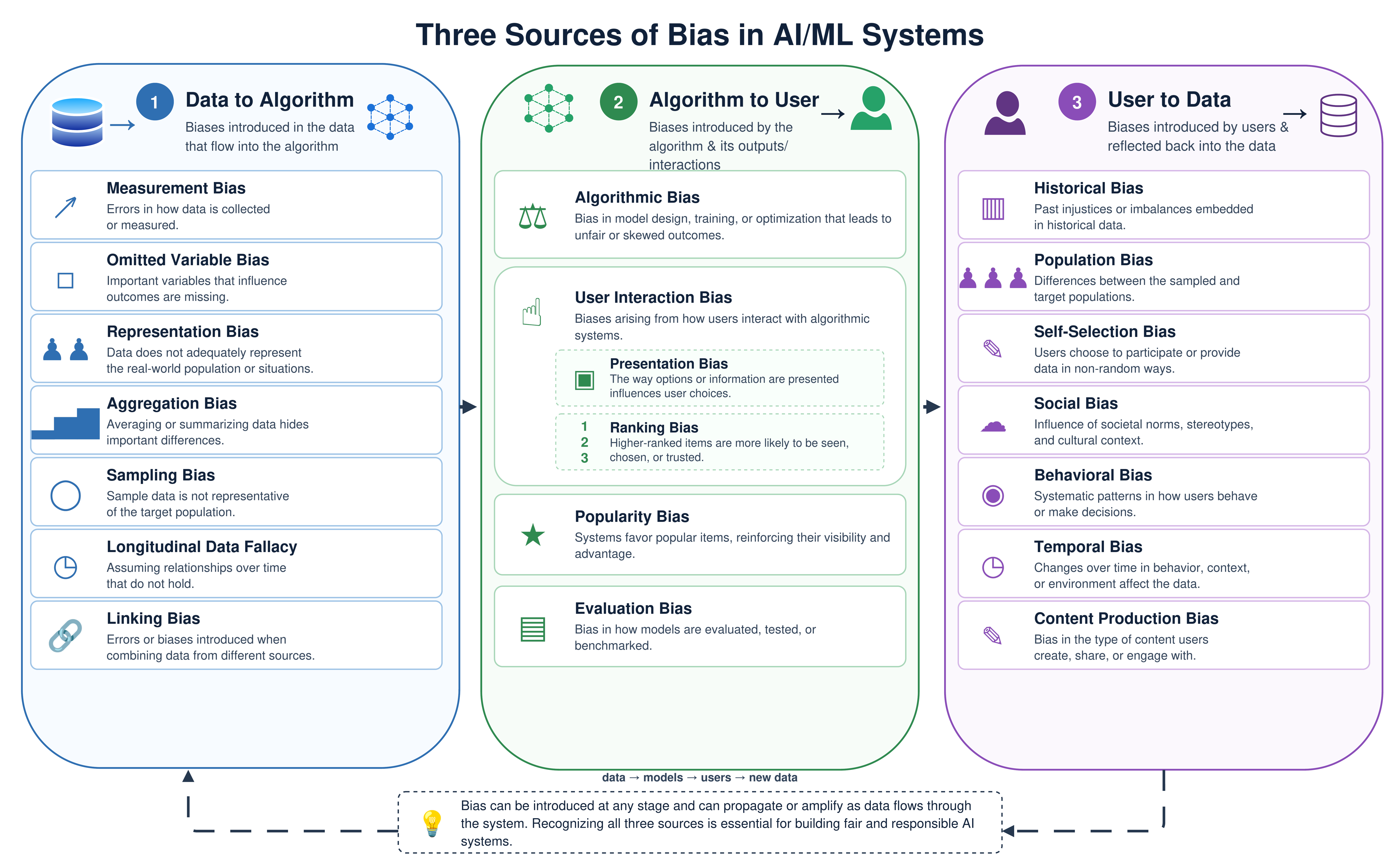}
    \caption{Types of Bias, adapted from \cite{mehrabi} }
    \label{fig:types_of_bias}
\end{figure}

\paragraph{Bias in Machine Learning: }The technical half of this discussion is grounded in the taxonomy proposed by Mehrabi et al.~\cite{mehrabi}, who organise sources of bias in AI systems into a data-algorithm-user feedback loop.
\emph{Data-to-algorithm} biases arise from how training data is collected, sampled and aggregated.
\emph{Algorithm-to-user} biases arise from model design and how outputs are presented to users.
\emph{User-to-data bias} arises from how user behaviour and historically embedded assumptions feed back into future data.
We adopt this structure, shown in Fig.~\ref{fig:types_of_bias}, as the organising frame for the remainder of this section and for the bias dimensions derived in Section~\ref{sec:vabiko_bias_mitigation}.

The taxonomy was developed primarily with contemporary, well-documented datasets and demographic fairness problems in mind (e.g. recidivism prediction, hiring, facial recognition).
It says comparatively little about the specific conditions in historical image archives: sparse or absent original metadata, image degradation, monochrome photography, and description authored under the assumptions of a colonial-era archivist rather than modern annotator.
Section~\ref{sec:vabiko_bias_dimensions} examines which categories of this taxonomy are most affected by these conditions, and Section~\ref{sec:vabiko_bias_mitigation} revisits the mitigation techniques associated with each in light of them.

\subsection{Bias Mitigation}
\label{sec:rw_bias_mitigation}
The following overview summarises bias mitigation techniques developed and evaluated in the general ML fairness literature. 
Section~\ref{sec:vabiko_bias_mitigation} revisits each of these in light of the specific bias dimensions and data particularities identified in Section~\ref{sec:context_vabiko}, discussing how they are applicable to this context.

\paragraph{Data Augmentation.} One important step in developing a domain-specific model for metadata generation or classification is training it on historical images.
However, historical image collections often contain substantial imbalances across attributes such as gender and race.
Creating a more balanced training dataset can help ensure that these attributes are represented more equitably and reduce the risk that certain groups become less visible in model predictions or retrieval results~\cite{mikolajczyk2018data,fahse2021managing}.
Data augmentation can support this process by increasing the diversity of under-represented classes.
Common techniques include image flipping, rotation, reflection, and transformations of colour, brightness, or contrast~\cite{mikolajczyk2018data}.
Such techniques are widely used in domains including medical imaging and speech recognition.
Although data augmentation has also been applied to cultural heritage collections, its use to address demographic imbalance in historical black-and-white imagery remains relatively limited~\cite{obeso2017architectural}.
One such study, conducted on the Architectural Heritage Elements (AHE) dataset and the Swedish World Culture Museum (WCM) dataset, applied noise injection, image flipping, and colour jittering as augmentation techniques and showed an increase in the performance of the classifier models.
Noise injection yielded the largest improvement, closely followed by colour jittering~\cite{foka2025}.

\paragraph{Adversarial Debiasing.}Adversarial debiasing is an in-processing technique in which a primary model learns the target task while an adversarial model attempts to predict a sensitive attribute from the primary model's learnt representation. 
The primary model is trained to jointly minimise its task loss and maximise the adversary's error, discouraging it from encoding or relying on that attribute~\cite{zhang2018mitigating,berg-etal-2022-prompt,wang2019balanced,lemoine2018mitigating}.
For example, in a historical image collection, a model trained to retrieve or classify images depicting people could be paired with an adversary that attempts to predict gender from the model's learnt representations, discouraging the model from associating the general concept of "person" more strongly with men than with women~\cite{lemoine2018mitigating}.

\section{VaBiKo Project: Context \& Use Cases}
\label{sec:context_vabiko}
\subsection{Project Goals}
\label{sec:vabiko_goals}

The VABiKo project deals with the photographs held by the Universitätsbibliothek J.\,C.\ Senckenberg (UB JCS) in Frankfurt am Main, collected between 1887 and 1943 by the Deutsche Kolonialgesellschaft (DKG) and its successor organisations.
The library also holds at least 20 further successor and other smaller collections from colonial contexts that arrived after 1995.
Its main goal is to improve access to the DKG holdings for scholars and a broader public.
Since usability is limited by incomplete metadata, the project pursues semi-automatic content indexing, using computer-vision methods to generate metadata that experts subsequently review and correct.
These components are to be adapted to the particularities of this material rather than adopted off the shelf, and are developed and evaluated for images of Persons of Colour from sub-Saharan Africa and Oceania (not only from former German colonies).
Finally, the project aims to balance openness with sensitive handling of material that may depict violence or carry terms offensive to the communities and countries of origin.

\subsection{Data Characteristics}
\label{sec:vabiko_data}

The photographs in question are predominantly glass-plate negatives and positive slides for presentations, alongside prints and photographic film, digitised through DFG-funded microfilming in the 1990s and held since 2021 in UB JCS's Visual Library system.
Cataloguing was performed by librarians in a largely content-descriptive fashion, since resources to research photographer, subject, place, or date against their historical context were not available.
The images may depict colonial violence, nudity, and displays of power, and reflect the hegemonic structures and racist stereotypes of their production context~\cite{risseewerth2024vabiko}; the metadata may also do so.
Despite these limitations, the collection has been openly accessible since 1999 and is actively used in publications, exhibitions, and teaching, in Germany and in the countries and communities of origin.
The term coverage of the full VABiKo metadata export of 62,519 records is summarised in Table~\ref{tab:metadata_coverage}.
Where present, content keywords are drawn from a controlled vocabulary developed for this collection~\cite{wolckerenk1995weiterbearbeitung}, organised as a hierarchy of up to six levels.

Where individuals are named, they are mostly colonial officials and political figures (Carl Peters, Hermann von Wissmann, Bismarck, Hitler).
The colonised population is overwhelmingly not named: 22.0\% of records ethnically ascribed as European carry a named individual, against 3.5\% of those ascribed African.
Most of the latter are titled rulers (e.g. Samuel Maharero, Hendrik Witbooi) or descriptions rather than names (``2 plantation workers,'' ``chief's daughter of Ovamboland'').

\begin{table}[htbp]
    \centering
    \caption{Term coverage across the VABiKo metadata (n = 62{,}519 records).}
    \label{tab:metadata_coverage}
    \begin{tabular}{lrl}
        \toprule
        Field & Coverage & Notes \\
        \midrule
        Title & 100.0\% & often only a shelfmark number \\
        Content keywords & 61.4\% & 8{,}124 unique terms; 6.7 per record on average \\
        Subject location & 60.9\% & 2{,}211 unique terms; 4.9 per record on average \\
        Text on envelope & 67.2\%  & transcribed, unindexed inscriptions \\
        Text on image & 44.9\% & transcribed, unindexed inscriptions \\
        Photographer & 33.8\% & \\
        Creation date (start year) & 23.0\% & \\
        Depicted person & 3.5\% & 1{,}272 unique named individuals \\
        \bottomrule
    \end{tabular}
\end{table}

\subsection{Use Cases for Search in Historical Image Collections}
\label{sec:vabiko_use_cases}

\subsubsection{Search for Location and Person Entities: }
This use case allows a query to match every name under which the same place or person is recorded, regardless of which term the searcher or the original cataloguer used. For example, a search for ``Deutsch-Ostafrika'' should also retrieve records tagged only with a later name, such as ``Tanzania'' or ``DOA'', and a search for ``Emin Pascha'' should retrieve records tagged under his birth name, ``Schnitzer, Eduard''.

\paragraph{Technical Solution.}
This requires an NER model to identify location and person mentions, and an entity-linking model, trained on mention-to-entry pairs, to resolve each mention and the query onto the same knowledge-base entry regardless of which name form is used.

\subsubsection{Domain-Expert Search by Content Keyword: }
\begin{figure}[htbp]
    \centering
    \begin{subfigure}[b]{0.32\textwidth}
        \includegraphics[width=\textwidth]{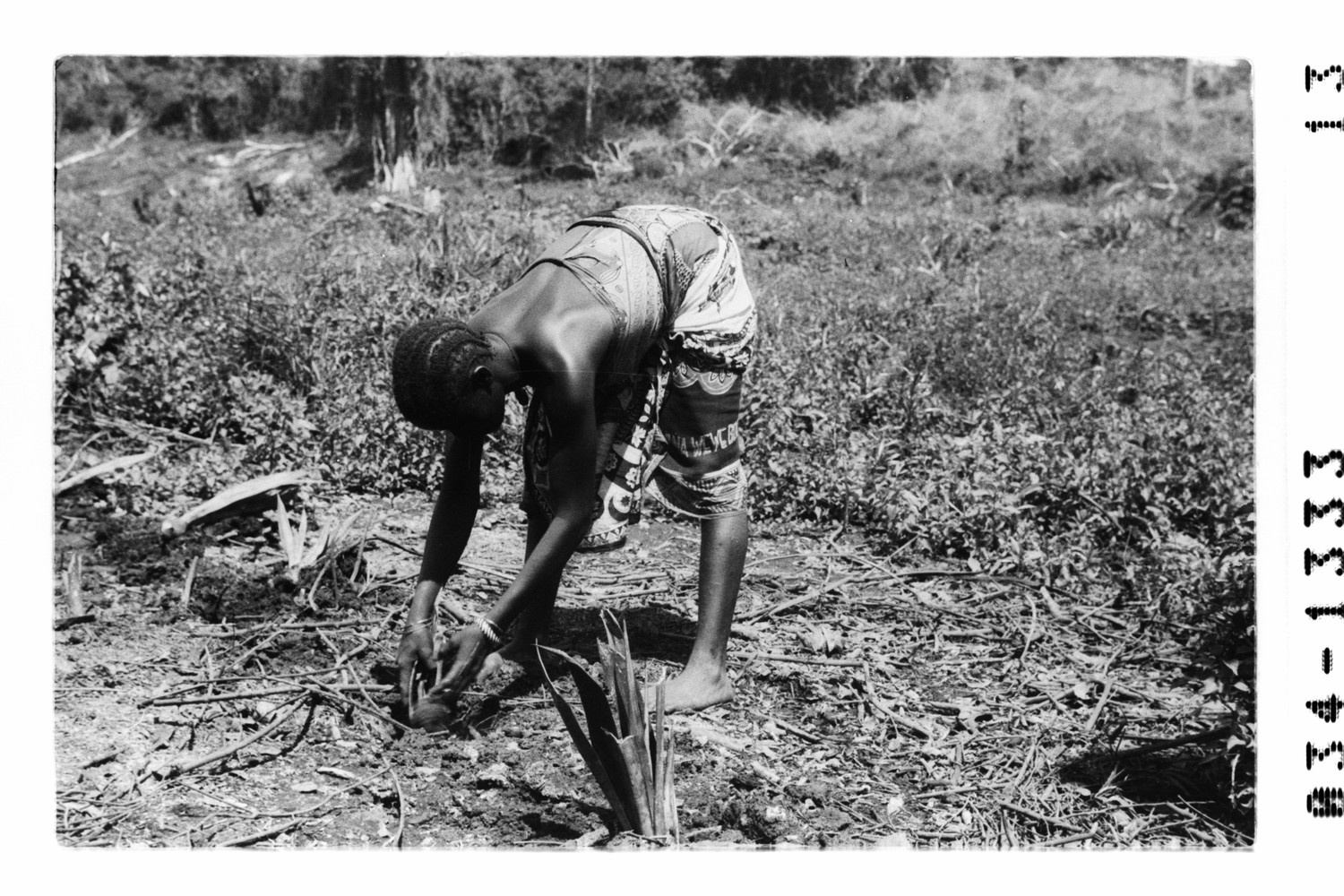}
        \caption{\textit{sisal agave}~\cite{vabiko_034133313}}
    \end{subfigure}
    \hfill
    \begin{subfigure}[b]{0.32\textwidth}
        \includegraphics[width=\textwidth]{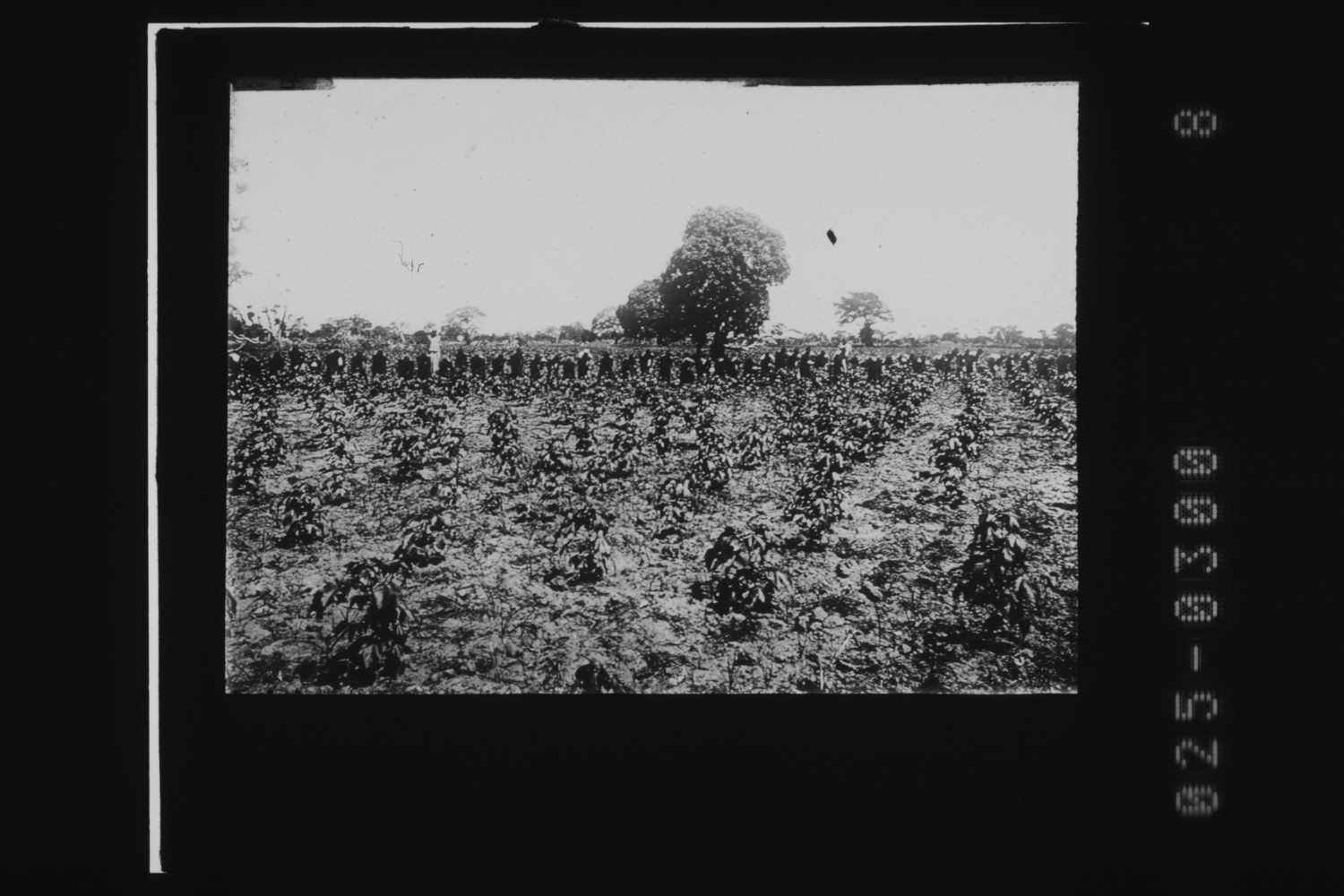}
        \caption{\textit{cotton plant}~\cite{vabiko_025030008}}
    \end{subfigure}
    \hfill
    \begin{subfigure}[b]{0.32\textwidth}
        \includegraphics[width=\textwidth]{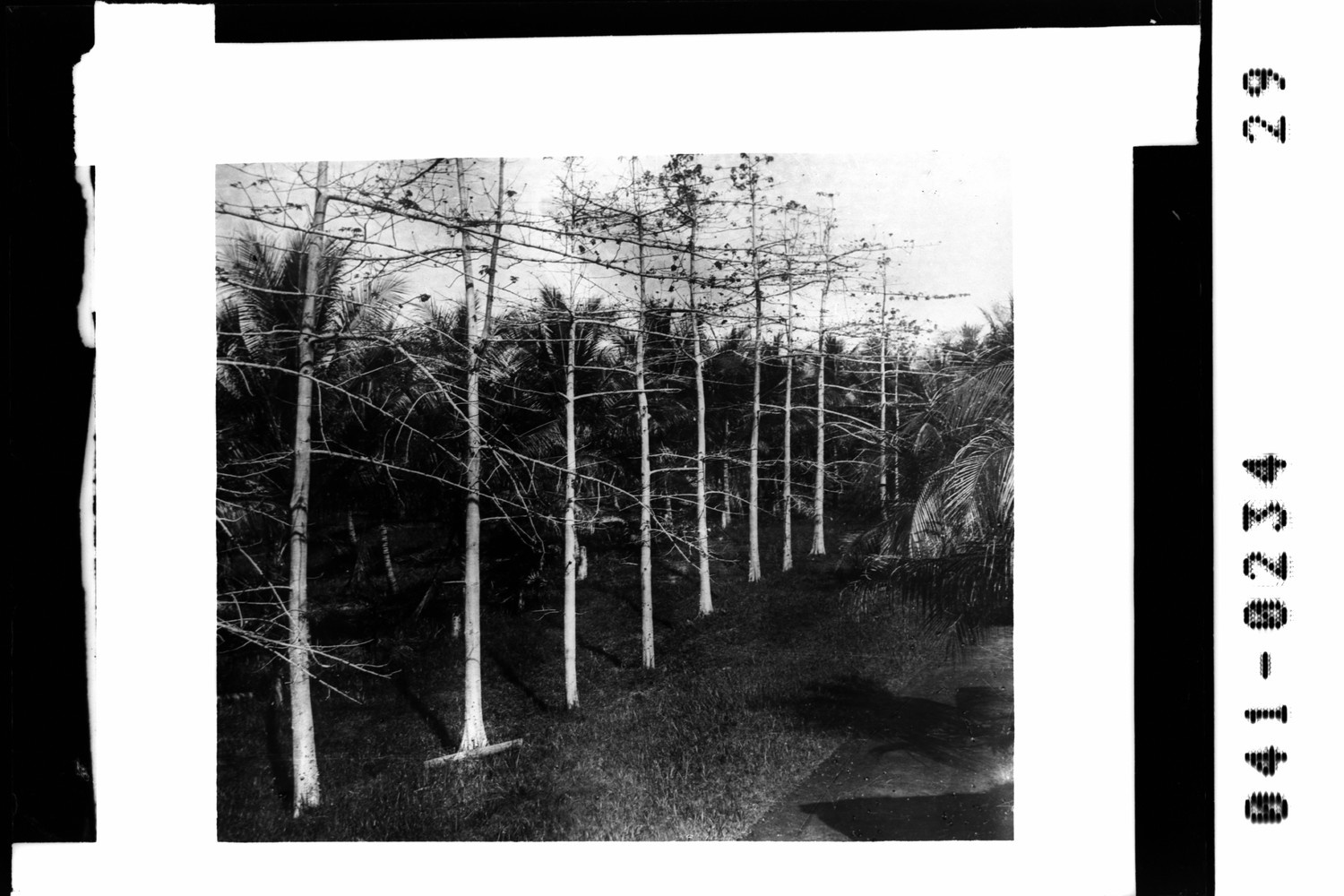}
        \caption{\textit{kapok tree}~\cite{vabiko_041023429}}
    \end{subfigure}
    \caption{Three images tagged with keywords under the category \textit{fibre-yielding plants}.}
    \label{fig:keyword_examples}
\end{figure}

The controlled vocabulary used for this project~\cite{wolckerenk1995weiterbearbeitung} is organised as a hierarchy of up to six levels, and each image is tagged by domain experts with its most specific applicable term. For example, the first record in Figure~\ref{fig:keyword_examples} is tagged \textit{sisal agave}; its only other label is a shelfmark number and the tag is the only way to know what the indistinct plant shown actually is. Since \textit{sisal agave} is a leaf term under the parent \textit{fibre-yielding plants}, that parent term also retrieves the other two records in the figure, cotton and kapok, even though the images are not visually similar.

\paragraph{Technical Solution.}
This requires a model, trained on images paired with their correct keywords, that predicts content keywords for the majority of records to which the controlled vocabulary was never applied.

\subsubsection{Free-Text Search Based on Image Content: }
This use case matches free-text queries against an image's content even when no keyword captures it. For example, one record tagged only \textit{hospitality trade} and \textit{colonial style} shows a building with a sign reading \textit{Hotel Rümmler}; a VLM caption (Qwen3-VL-8B-Instruct) describes the scene and reproduces this name, which appears nowhere in the record's metadata, so only a query against the image or its caption can retrieve it.

\begin{figure}[htbp]
    \centering
    \begin{subfigure}[b]{0.28\textwidth}
        \includegraphics[width=\textwidth]{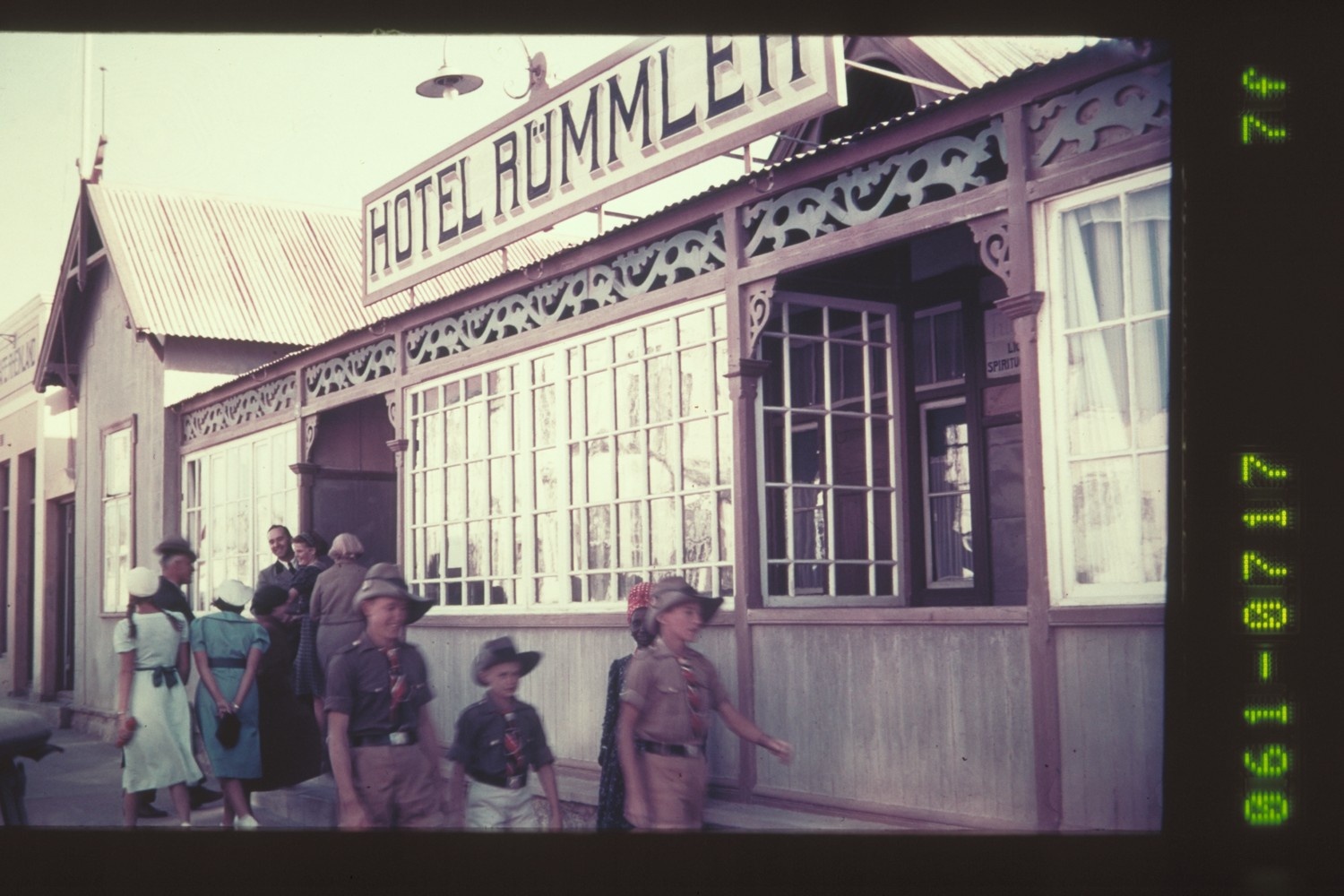}
        \caption{Tagged: hospitality trade, colonial style~\cite{vabiko_061071707}}
    \end{subfigure}
    \hfill
    \begin{subfigure}[b]{0.28\textwidth}
        \includegraphics[width=\textwidth]{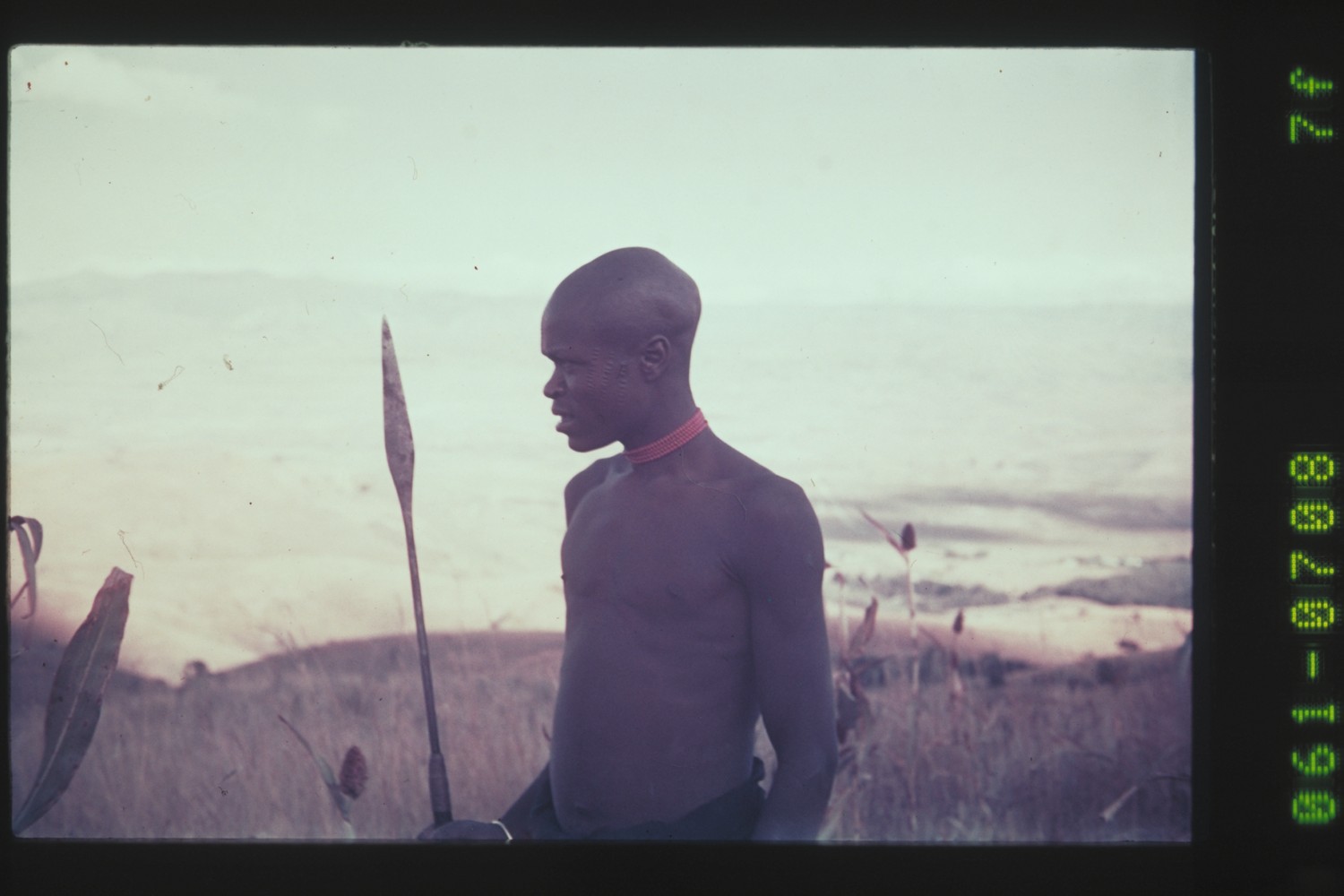}
        \caption{Tagged: man, spear, weapons, population~\cite{vabiko_061070807}}
    \end{subfigure}
    \caption{Two records whose existing tags (below each image) do not capture information that VLM captions can provide.}
    \label{fig:content_only_examples}
\end{figure}

\paragraph{Technical Solution.}
This is built on a VLM that captions every image, against which free-text queries are matched lexically or by embedding similarity, or on a shared image-text embedding model such as CLIP that ranks images directly by similarity to the query.

\subsubsection{Near-Duplicate Linking: }
\begin{figure}[htbp]
    \centering
    \begin{subfigure}[b]{0.32\textwidth}
        \includegraphics[width=\textwidth]{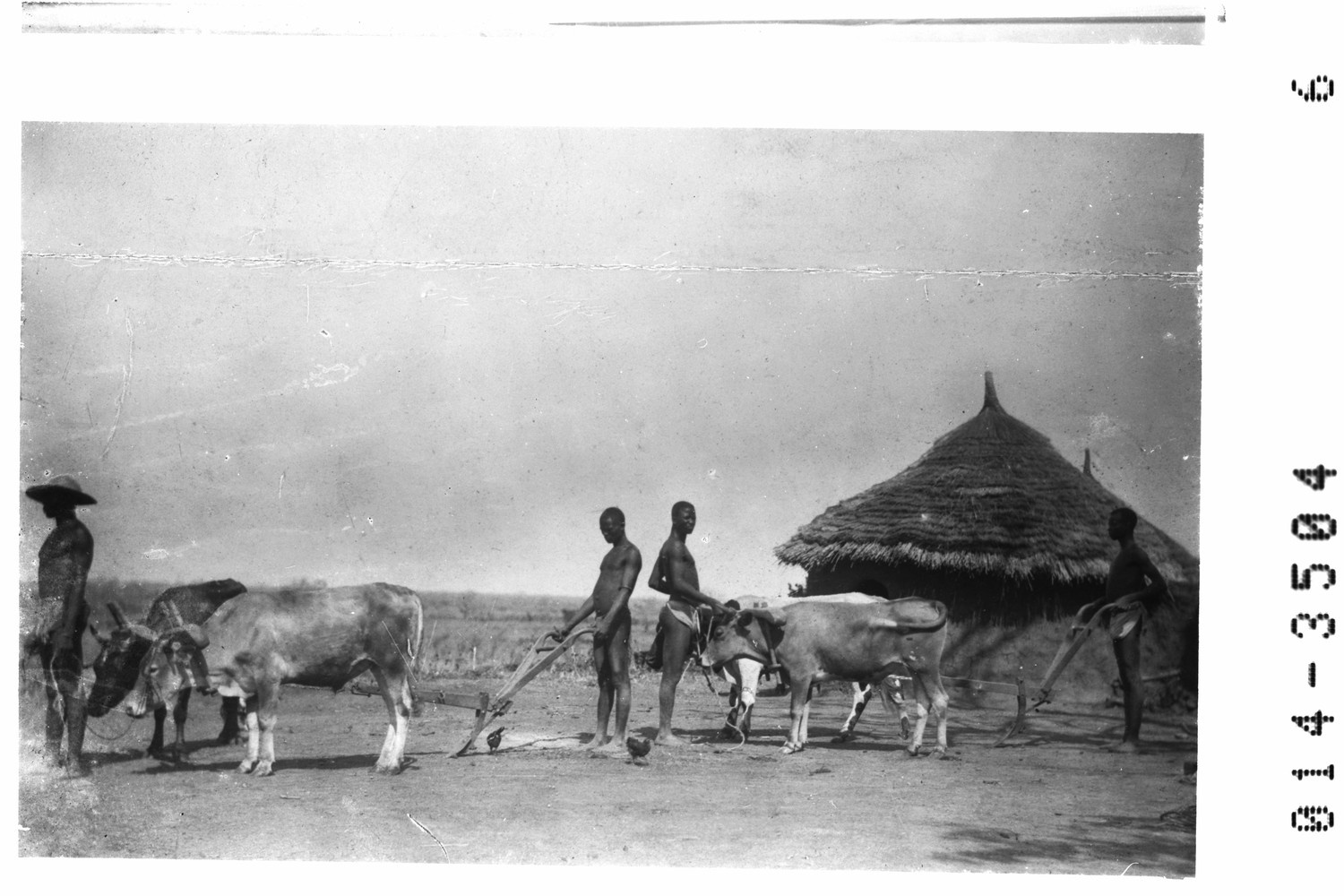}
        \caption{014-3504-06~\cite{vabiko_014350406}}
    \end{subfigure}
    \hfill
    \begin{subfigure}[b]{0.32\textwidth}
        \includegraphics[width=\textwidth]{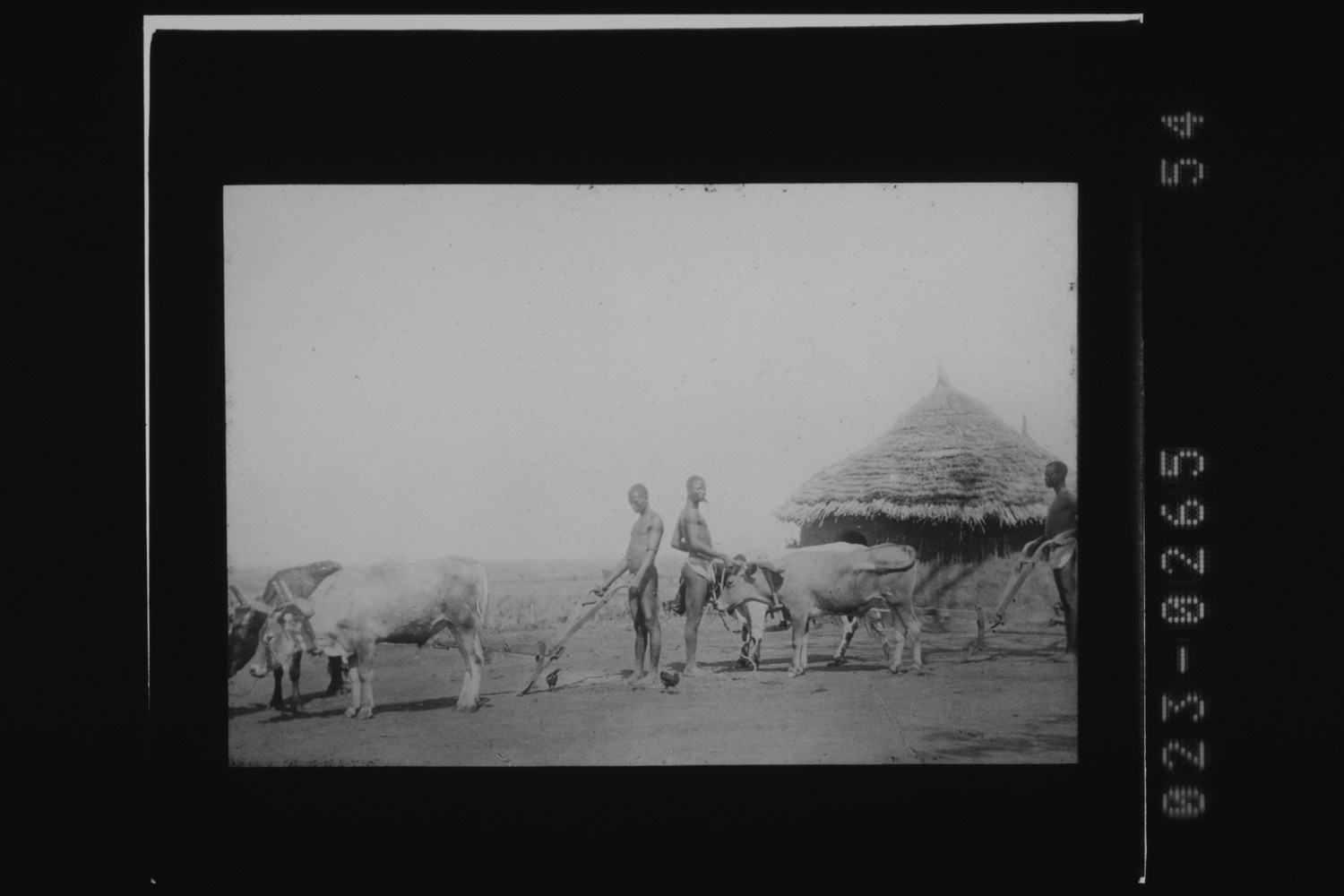}
        \caption{023-0265-54~\cite{vabiko_023026554}}
    \end{subfigure}
    \hfill
    \begin{subfigure}[b]{0.32\textwidth}
        \includegraphics[width=\textwidth]{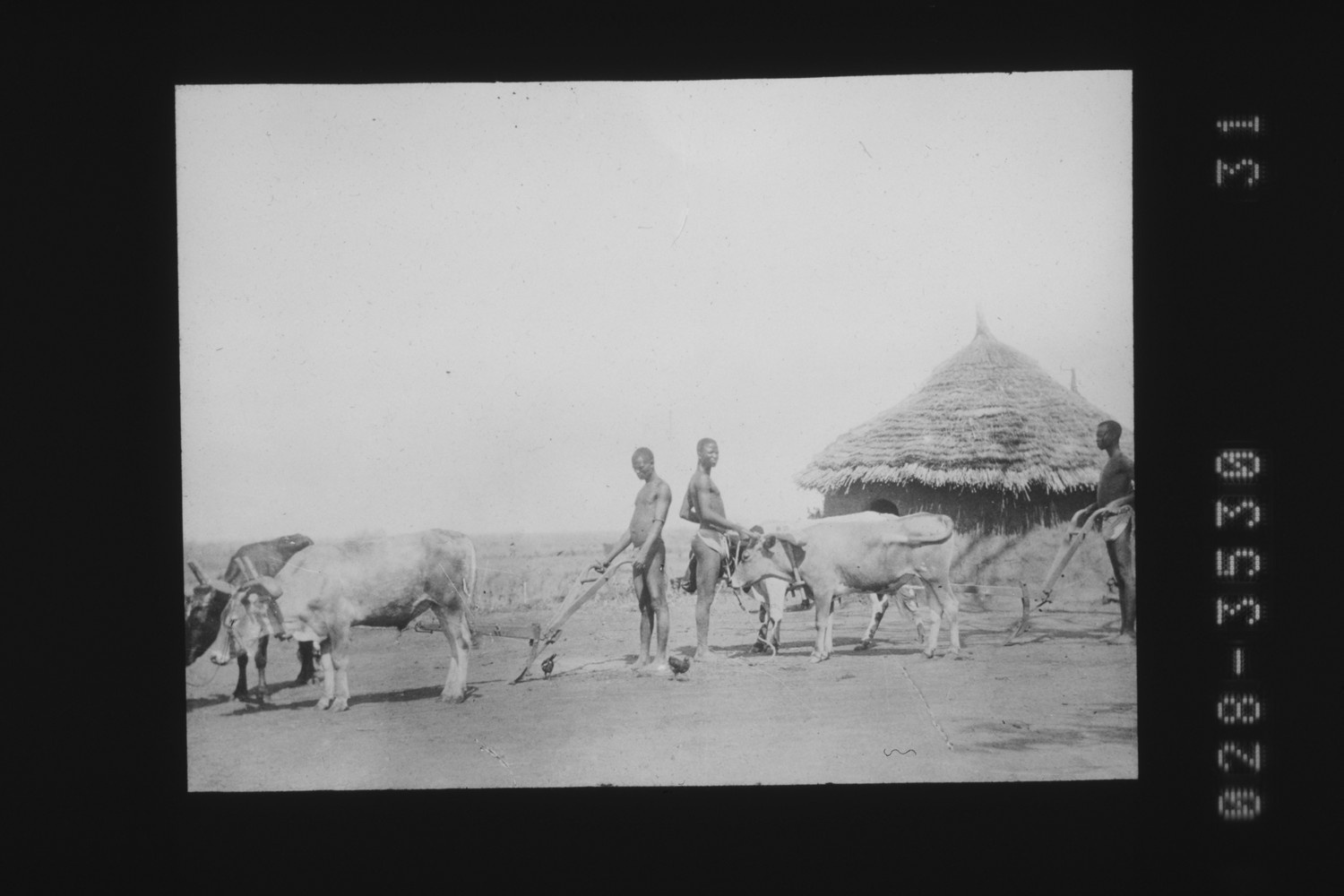}
        \caption{028-3530-31~\cite{vabiko_028353031}}
    \end{subfigure}
    \caption{A near-duplicate photograph held under three identifiers, with differing captions.}
    \label{fig:togo_duplicates}
\end{figure}

This use case links all copies and near-duplicate images across the corpus.
Figure~\ref{fig:togo_duplicates} shows one photograph held under three identifiers: a reproduction print captioned with the specific village and activity depicted, and two lantern slides that instead carry a series-level label from a numbered lecture-slide set. Linking such duplicates lets a historian trace not only reuse but also where an image-specific caption was replaced by a generic one.

\subsubsection{Query-by-Example Search: }
\begin{figure}[htbp]
    \centering
    \begin{subfigure}[b]{0.32\textwidth}
        \includegraphics[width=\textwidth]{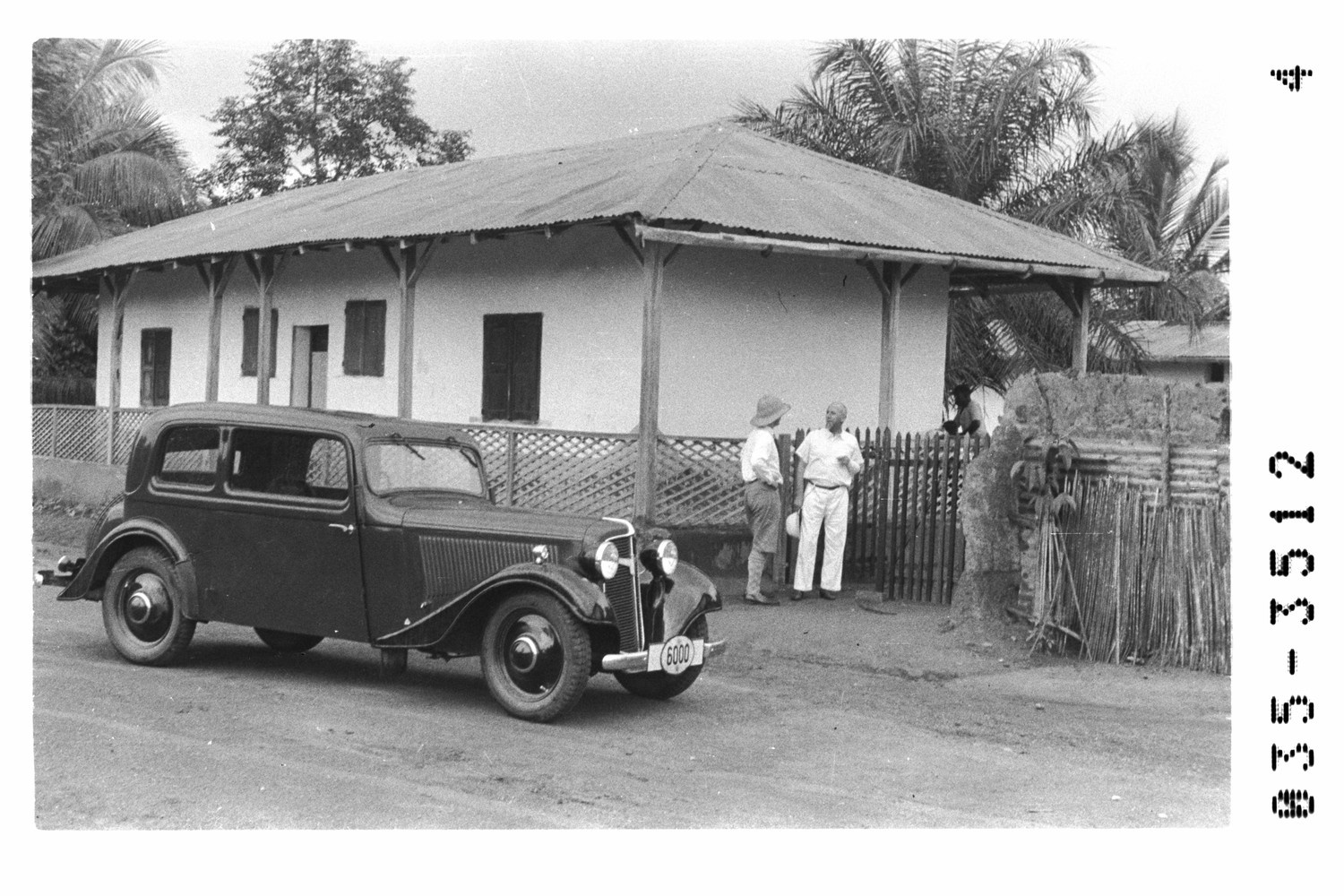}
        \caption{Tagged: car, stone house~\cite{vabiko_035351204}}
    \end{subfigure}
    \hfill
    \begin{subfigure}[b]{0.32\textwidth}
        \includegraphics[width=\textwidth]{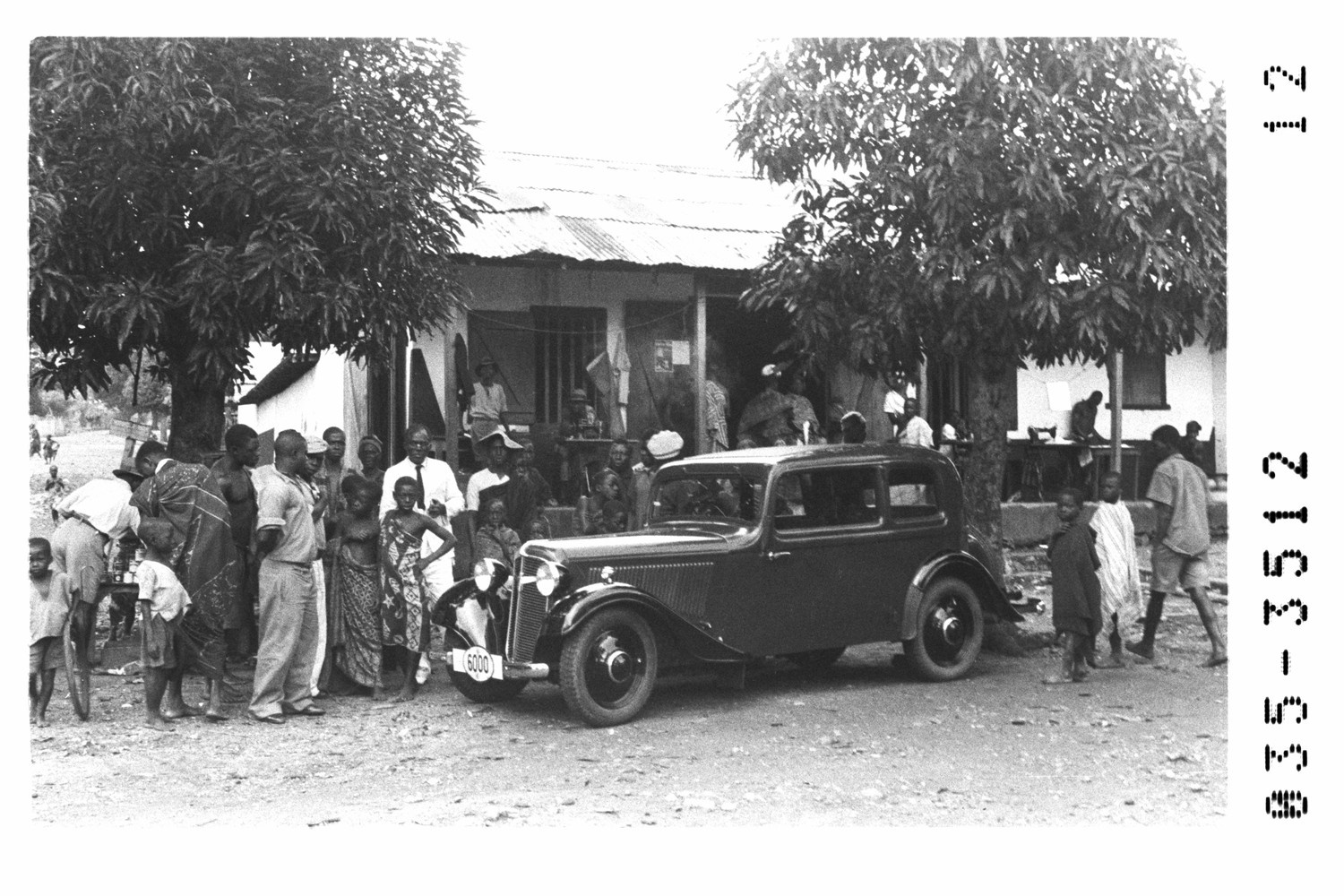}
        \caption{Tagged: street scene, traffic~\cite{vabiko_035351212}}
    \end{subfigure}
    \hfill
    \begin{subfigure}[b]{0.32\textwidth}
        \includegraphics[width=\textwidth]{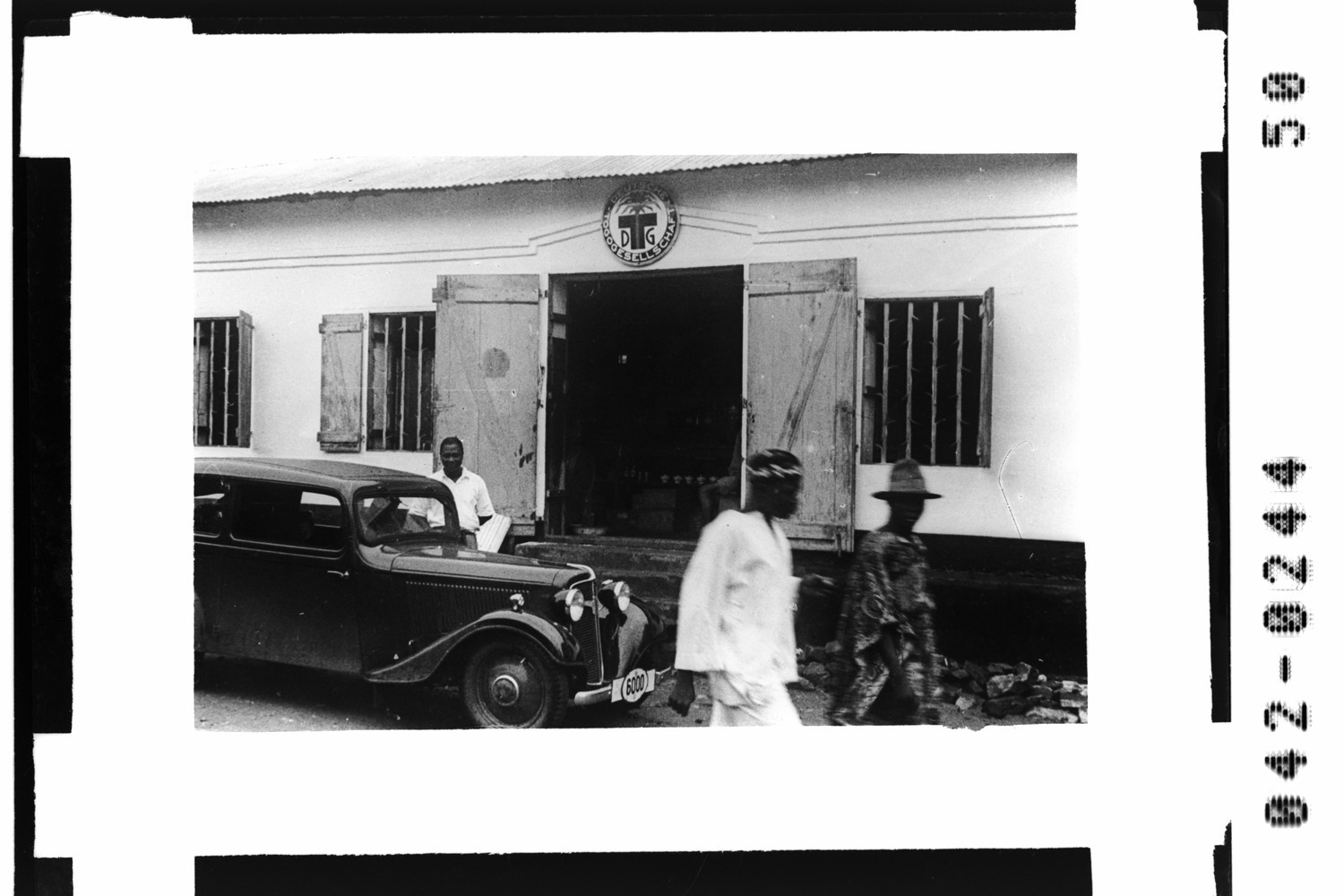}
        \caption{Tagged: licence plate, 6000~\cite{vabiko_042024450}}
    \end{subfigure}
    \caption{The same car, at three different buildings.}
    \label{fig:same_car}
\end{figure}

This use case retrieves related material by querying with an example image.
Figure~\ref{fig:same_car} shows three records of the same car, identifiable by its licence plate, photographed at three different buildings; only one record names the plate as a keyword, so a keyword search would surface just that one. Visual similarity ranks all three as mutual nearest neighbours (cosine similarity 0.81--0.85) regardless of which record names the car, letting a researcher reconstruct the connection from any one of them.

\paragraph{Technical Solution.}
These use cases, together with a third (similar-image suggestions, which pre-computes and statically displays each record's top-$k$ nearest neighbours), share a pre-trained visual encoder that maps every image onto an embedding stored in a similarity index. Near-duplicate linking clusters embeddings above a similarity threshold, while query-by-example returns the nearest neighbours of a user-supplied image. Near-duplicate detection typically needs a differently-tuned encoder, since it targets a stricter notion of similarity than the other two use cases.

\subsection{Relevant Bias Dimensions}
\label{sec:vabiko_bias_dimensions}

Table~\ref{tab:vabiko_bias_dimensions} summarises the bias dimensions documented for each of the tasks introduced above, together with the stage of the data-algorithm-user cycle at which each arises.
Entity linking is affected at both ends of who was documented, since colonised individuals are under-covered in reference knowledge bases and disambiguation models rank the remaining candidates by popularity.
Multi-label image classification is affected by historical bias and a long-tailed label distribution from the controlled vocabulary, representation bias from pre-training corpora and from image degradation, and measurement bias from predicting on the image alone.
VLM captioning inherits the skin-colour disparity of the captions it is trained on and amplifies it, which we label algorithmic bias.
Text-based image retrieval carries representation bias from CLIP-family embeddings trained on modern web data, and evaluation bias because the conventional fairness benchmarks are built from contemporary photographs.
Visual similarity search is affected by content production bias because near-duplicates here arise from reproduction prints and lantern slides rather than from the camera variation its benchmarks assume.

\begin{table}[htbp]
\centering
\caption{Bias dimensions documented for each task discussed in this section.}
\label{tab:vabiko_bias_dimensions}
\small
\begin{tabular}{|p{0.20\linewidth}|p{0.17\linewidth}|p{0.53\linewidth}|}
\hline
\multicolumn{1}{|c|}{\textbf{Task}} & \multicolumn{1}{c|}{\textbf{Bias Type}} & \multicolumn{1}{c|}{\textbf{Description}} \\
\hline
Entity Linking & \textbf{algorithm-to-user}\newline\textit{Popularity} & Disambiguation models rank candidates by entity popularity favouring individuals from overrepresented nationalities~\cite{rao2013entity}. \\
\hline
Entity Linking & \textbf{data-to-algorithm}\newline\textit{Representation} & Colonized individuals are under-covered in reference knowledge bases, limiting which entities are retrievable as candidates~\cite{yu2025demographic}. \\
\hline
Multi-Label Image Classification & \textbf{data-to-algorithm}\newline\textit{Sampling} & Long-tailed label distributions in the training data degrade performance on under-represented categories~\cite{alijani2025historyclip}. \\
\hline
Multi-Label Image Classification & \textbf{data-to-algorithm}\newline\textit{Representation} & Pre-training datasets skew toward American and European imagery, lowering accuracy on images from other locations~\cite{shankar2017geodiversity}. \\
\hline
Multi-Label Image Classification & \textbf{data-to-algorithm}\newline\textit{Measurement} & Predicting from the image alone omits context (location, date, purpose) that would improve accuracy~\cite{macaodha2019presence}. \\
\hline
Multi-Label Image Classification & \textbf{data-to-algorithm}\newline\textit{Representation} & Physical degradation from aging, transfers and digitisation differs from the born-digital images classifiers are trained on~\cite{pei2019degradation}. \\
\hline
Multi-Label Image Classification & \textbf{user-to-data}\newline\textit{Historical} & Controlled vocabularies encode the priorities of designers which flattening other perspectives~\cite{warburton2024quantifying}. \\
\hline
VLM Captioning & \textbf{algorithm-to-user}\newline\textit{Algorithmic} & Captioning systems amplify the racial disparities already present in the captions they are trained on~\cite{zhao2021racial}. \\
\hline
VLM Captioning & \textbf{user-to-data}\newline\textit{Historical} & Human-written captions vary in sentiment depending on the subject's skin colour~\cite{zhao2021racial,facklam2025describing}. \\
\hline
Text-Based Image Retrieval & \textbf{data-to-algorithm}\newline\textit{Representation} & CLIP-family embeddings trained on modern web data over-expose White and male subjects relative to the collection's demographics~\cite{wolfe2022markedness,awale2026examining}. \\
\hline
Text-Based Image Retrieval & \textbf{algorithm-to-user}\newline\textit{Evaluation} & Standard fairness benchmarks (FairFace, FACET) use contemporary photographs and miss biases specific to historical imagery~\cite{karkkainen2021fairface,gustafson2023facet}. \\
\hline
Visual Similarity Search & \textbf{user-to-data}\newline\textit{Content production} & Reproduction prints, lantern slides, and repeated photography differ from the camera/lighting variation that near-duplicate benchmarks are built from~\cite{net2024transductive}. \\
\hline
\end{tabular}
\end{table}

\section{Bias Mitigation Techniques for Search in Historical Images}
\label{sec:vabiko_bias_mitigation}
For each mitigation below we state which dimension of Table~\ref{tab:vabiko_bias_dimensions} it addresses and what the literature reports about it.
We group them by where they act: on the metadata and reference vocabularies, on the data a model is trained on, or on how a model learns or operates.

\subsection{Curatorial Interventions}
\label{sec:curatorial_interventions}

\emph{Contextualising contentious terms rather than deleting them} addresses the historical bias of a controlled vocabulary~\cite{menis-mastromichalakis-etal-2025-dont}.
The original description stays in place as evidence of how a collection's creators classified their material, and explanations and alternative terms are added alongside it.
This changes individual terms but not the shape of the vocabulary.
Where a branch flattens several concepts into one term, that form of historical bias remains~\cite{warburton2024quantifying}.
\emph{Knowledge-graph enrichment} addresses the under-coverage of colonised individuals in reference knowledge bases~\cite{angelis2024chekg,kraft-usbeck-2022-lifecycle}.
Entities are extracted from further archival sources, alternative and multilingual name forms are linked, and provenance and uncertainty are recorded, with experts validating what is added.

\subsection{Data-side Interventions}
\label{sec:data-side_interventions}

\emph{Data augmentation} (Section~\ref{sec:rw_bias_mitigation}) addresses long-tailed label distributions and the American and European skew of pretraining corpora.
It synthesises extra examples for the classes that have few.
Two of the operations reported as most effective on cultural-heritage collections, colour jittering and colour transformation, may not work on monochrome material.
\emph{Adapting the model to the imbalance instead of resampling around it} is the closer fit where the imbalance comes from the vocabulary rather than from sampling.
HistoryCLIP does this for retrieval over long-tailed archival data~\cite{alijani2025historyclip}.
\emph{Training on artificially degraded images} addresses the degradation dimension.
The EyCon project did this by imitating the halftone reproduction of its press sources~\cite{dentler2024sensitivity}.
The degradation has to be modelled on the carrier and digitisation chain of the collection at hand.
\emph{Conditioning prediction on non-visual metadata} addresses measurement bias~\cite{macaodha2019presence}, and is feasible here because location and date are often already recorded.
\emph{Hierarchical multi-label learning} addresses the long tail.
Tying a leaf prediction to its ancestors lets rare terms share training signal with their parents~\cite{wehrmann2018hierarchical,zhang2022usealllabels}.

\subsection{Algorithm-side Interventions}
\label{sec:algorithm-side_interventions}

\emph{Ranking candidates by the context around a mention rather than by how well known they are} addresses the popularity dimension of entity linking~\cite{rao2013entity,ayoola-etal-2022-improving}.
Popularity is usually a good guess, but it always favours the better-documented entity.
Such systems can also answer that no entity matches, which matters when the right one is missing from the knowledge base~\cite{arora-etal-2024-contrastive}.
\emph{Adversarial debiasing} is used when a model encodes an attribute such as gender or skin colour in its representation and allows it to shape the results~\cite{berg-etal-2022-prompt}.
In text-based retrieval this shows up as a query that names no group returning mostly White and male subjects~\cite{wolfe2022markedness}.
There is evidence that it does not need to transfer to historical material.
Evaluated on one collection of colonial-context photographs, DebiasCLIP retrieved images of women less accurately than the unmitigated baseline.
Queried for a \textit{White person}, it returned no White individuals and further over-exposed Black subjects~\cite{awale2026examining}.
\emph{Annotation-free attribute neutralisation}~\cite{hirota2025saner} and \emph{joint debiasing of both modalities}~\cite{Zhang_2025_CVPR} address the same dimension.
Neither needs labelled demographic groups, which colonial collections rarely have, but neither has been evaluated on historical material.
\emph{Transductive self-supervised fine-tuning on the target collection} addresses the near-duplicate dimension~\cite{net2024transductive}.
Unlike the methods above, it rests on an assumption archival deployment satisfies: the target set is known in advance.
It therefore needs neither labels nor an external benchmark.
\emph{Rewriting captions using a model trained on synthetically biased examples} addresses both captioning dimensions~\cite{hirota2023libra}.
It is built for gender in contemporary photographs, whereas the disparity documented for historical material is racial~\cite{Hirota_2022_CVPR}.
\emph{Domain-specific evaluation data} conditions all of the above.
FairFace and FACET are built from contemporary photographs~\cite{karkkainen2021fairface,gustafson2023facet}, so a mitigation can pass on them and still fail on historical imagery, as DebiasCLIP did.
Outside text-based retrieval~\cite{awale2026examining}, no comparable resource exists.

\section{Conclusion}
This paper examines bias in AI-supported retrieval for colonial-context photographic archives from humanities and machine-learning perspectives. Grounded in the VaBiKo collection, we map how historical, representational, metadata, algorithmic, ranking, and evaluation biases arise at different stages of the retrieval pipeline.
The analysis shows that incomplete descriptions, uneven representation, rare entities, and black-and-white imagery can cause bias.
It reviews the standard mitigation techniques that have been deployed in similar contexts to address these biases.

What helps with dealing with bias is a series of small corrections at different points, and setting tasks, bias types, and mitigations out together shows which of those points a given project can reach.
This paper's methodology works from use cases to the tasks they require, from those tasks to the documented bias types, and from those to mitigations reported in the literature.
This progression is reusable -- other GLAM institutions can apply the same steps to their own collections.
It turns a general concern about bias into a set of concrete decisions to be taken before a system is built, and it makes visible where no mitigation, or no suitable evaluation data, is available yet.

\label{sec:conclusion}
%
%
%
 \bibliographystyle{splncs04}
 \bibliography{references}

\end{document}